\documentclass[twocolumn,superscriptaddress,prl]{revtex4}
\usepackage{graphicx,bm}
\usepackage{graphicx,bm}
\usepackage{subfigure}
\usepackage{amssymb}
\usepackage{wasysym}

\newcommand{\la}{\langle}
\newcommand{\ra}{\rangle}

\newcommand{\lp}{\left(}
\newcommand{\rp}{\right)}

\newcommand{\be}{\begin{equation}}
\newcommand{\ee}{\end{equation}}
\newcommand{\bea}{\begin{eqnarray}}
\newcommand{\eea}{\end{eqnarray}}

\renewcommand{\phi}{\varphi}
\renewcommand{\epsilon}{\varepsilon}

\begin{document}

\title{Landau Level Collapse in Gated Graphene Structures}
\author{Nan Gu}
\affiliation{Department of Physics,
Massachusetts Institute of Technology, 77 Massachusetts Ave,
Cambridge, MA 02139}
\author{Mark Rudner}
\affiliation{Department of Physics,
Harvard University, 17 Oxford St.,
Cambridge, MA 02138}
\author{Andrea Young}
\affiliation{Department of Physics, Columbia University, New York, NY10027}
\author{Philip Kim}
\affiliation{Department of Physics, Columbia University, New York, NY10027}
\author{Leonid Levitov}
\affiliation{Department of Physics,
Massachusetts Institute of Technology, 77 Massachusetts Ave,
Cambridge, MA 02139}

\begin{abstract}
We describe a new regime of magnetotransport in two dimensional electron systems in the presence of a narrow potential barrier imposed by external gates.
In such systems, the Landau level states, confined to the barrier region in strong magnetic fields, undergo a deconfinement transition as the field is lowered. We present transport measurments showing Shubnikov-de Haas (SdH) oscillations which, in the unipolar regime, abruptly disappear when the strength of the magnetic field is reduced below a certain critical value. This behavior is explained by a semiclassical analysis of the transformation of closed cyclotron orbits into open, deconfined trajectories. Comparison to SdH-type resonances in the local density of states is presented.
\end{abstract}

\maketitle

Electron cyclotron motion constrained by crystal boundaries displays many interesting phenomena, such as skipping orbits and electron focusing, which have yielded a wealth of information on scattering mechanisms in solids \cite{Khaikin69,Tsoi99}.
Since the 1980s, many groups have investigated electron transport in semiconducting two-dimensional electron systems (2DES), where gate-induced spatially varying electric fields can be used to alter cyclotron motion. A variety of interesting phenomena were explored in these systems, including quenching of the quantum Hall effect \cite{Roukes,Ford}, Weiss oscillations due to commensurability between cyclotron orbits and a periodic grating \cite{Weiss oscillations}, pinball-like dynamics in 2D arrays of scatterers \cite{Weiss91}, and coherent electron focusing \cite{van Houten89}.

The experimental realization of graphene \cite{Novoselov06}, a new high-mobility electron system, affords new opportunities to explore effects that were previously inaccessible. In particular, attempts to induce sharp potential barriers in III-V semiconductor quantum well structures have been limited by the depth at which the 2DES is buried---typically about 100nm below the surface \cite{shayegan94}. In contrast, electronic states in graphene, a truly two-dimensional material, are fully exposed and thus allow for potential modulation on $\sim 10\,{\rm nm}$
length scales using small local gates and thin dielectric layers \cite{Huard07,Williams07,Ozyilmaz07,Young09}.
Significantly, such length scales can be comparable to the magnetic length $\ell_B=\sqrt{\hbar c/eB}$, which characterizes electronic states in quantizing magnetic fields. In this Letter we focus on one such phenomenon, the transformation of the discrete Landau level spectrum to a continuum of extended states in the presence of a static electric field.  

\begin{figure}
\includegraphics[width=3.4in]{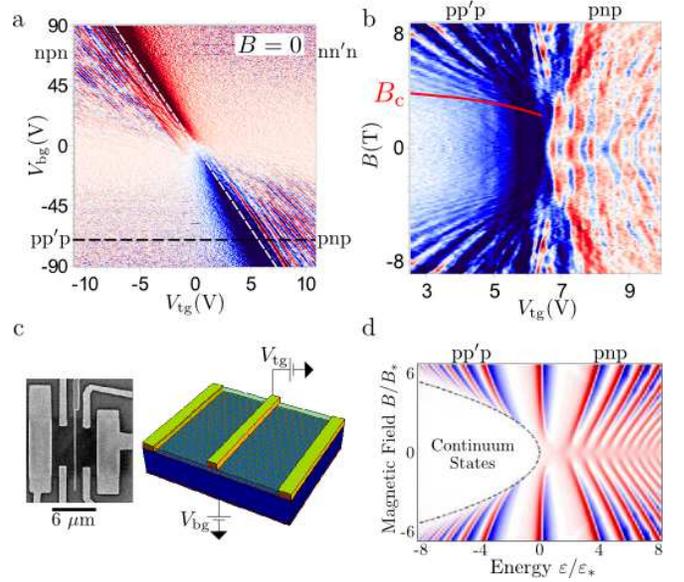}
 \caption[]{(a) Differentiated conductance, $dG/dV_{\rm tg}$, of a narrow top gate graphene device, pictured in (c). Fabry-Perot (FP) 
oscillations appear in the presence of confining pn junctions. (b) $dG/dV_{\rm tg}$ as a function of $B$ and $V_{\rm tg}$.  Shubnikov-de Haas (SdH) oscillations are observed at high $B$.  The fan-like SdH pattern is altered  by the barrier: in the ${\rm pp'p}$ region it curves, weakens, and 
is washed out at fields $|B|\lesssim B_c$, Eq.(\ref{eq:critical_field_from_n(x)}), while in the pnp region a crossover to FP oscillations occurs. Data shown correspond to $V_{\rm bg}=-70{\rm V}$ [dashed line in (a)]. (c) Top gated graphene device micrograph and schematic; top gate width is $\sim{\rm 16nm}$. (d) Local density of states (DOS) in the middle of parabolic barrier, obtained numerically from Eqs.(\ref{eq:DOS}),(\ref{QuantumH}).
The derivative with respect to energy, $dN/d\epsilon$, which corresponds to the measured quantity $dG/dV_{\rm tg}$, is shown.
Dashed parabola marks the critical field,
Eq.(\ref{eq:Bc_quasiclassical}). Oscillations in the DOS modulate the rate of scattering by disorder, resulting in the SdH effect \cite{Abrikosov72}. 
}
\label{expt}
\vspace{-6mm}
\end{figure}

The behavior which will be of interest for us is illustrated by a toy model involving the Landau levels of a massive charged particle in the presence of an inverted parabolic potential $U(x)=-a x^2$. Competition between the repulsive potential and magnetic confinement gives rise to a modified harmonic oscillator spectrum
\be\label{eq:inverted_parabola}
\epsilon_n(p_y) = \frac{\hbar e}{m}\sqrt{B^2 - B_c^2} \left(n + 1/2\right) - \frac{2ap_y^2}{e^2(B^2-B_c^2)}
\ee
for $B>B_c$, where $m$ is the particle mass, $p_y$ is the $y$ component of momentum, and $B_c = \sqrt{2ma}/e$ is the critical magnetic field strength. For strong magnetic field, $B>B_c$, 
the spectrum consists of discrete (but dispersive) energy bands indexed by an integer $n$, whereas for $B\le B_c$ the spectrum is continuous even for fixed $p_y$.  This behavior can be understood quasiclassically in terms of transformation of closed cyclotron orbits into open orbits, which occurs when the Lorentz force is overwhelmed by the repulsive barrier potential.

For massless Dirac charge carriers in single-layer graphene, 
Landau levels subject to a linear potential $U(x) = -eEx$ manifest an analogous collapse of the discrete spectrum below the critical field $B_c=E/v_F$ \cite{Lukose2007}:
\be\label{eq:linear_U(x)}
\epsilon_n(p_y) =\pm v_F \sqrt{2n\hbar e B}\lp 1-\beta^2\rp^{3/4} - \beta  v_F p_y
,
\ee
where $n=0,1,2...$ and $\beta =  E/v_FB$. The transition at $B=B_c$ can be linked to the classical dynamics of a massless particle, characterized by closed orbits at $B>B_c$ and open trajectories at $B<B_c$ \cite{Shytov09}.

A simple but intuitive picture of the spectrum (\ref{eq:linear_U(x)})
can be obtained from the Bohr-Sommerfeld quantization condition 
\bea\label{eq:BS_quantization_general}
&&\int_{x_1}^{x_2}p_x(x)dx=\pi\hbar(n + 1/2-\gamma)
,\quad
\\\label{eq:BS_quantization_general_2}
&&p_x(x)=\sqrt{(\epsilon-U(x))^2/ v_F^2 - (p_y - eBx)^2}
,
\eea
where $x_1$ and $x_2$ are the turning points, $p_y$ is the conserved momentum parallel to the barrier, and the Berry phase contribution $\gamma$ is $1/2$ for Dirac fermions.
For linear $U(x)$, this gives the Landau level spectrum (\ref{eq:linear_U(x)}) for $B>B_c$. As $B$ approaches $B_c$, one of the turning points moves to infinity, indicating a transformation of closed orbits into open trajectories. 

To realize the collapse of Landau levels in an electron system, several conditions must be met. First, it must be possible to create a potential barrier that is steep on the scale of the cyclotron orbit radius. Second, the system must be ballistic on this length scale, in order to suppress the broadening of Landau levels due to disorder. Graphene, which is a truly two-dimensional material with high electron mobility, fulfills both conditions. Crucially, as demonstrated by the recent observation of Fabry-Perot (FP) oscillations in gated graphene structures \cite{Young09}, transport can remain ballistic even in the presence of a gate-induced barrier. Thus graphene is an ideal system for studying the Landau level collapse.

Transport data taken from a locally gated device similar to that described in Ref.\cite{Young09} are shown in Fig.\ref{expt}. Graphene was prepared via mechanical exfoliation and contacted using electron beam lithography before being coated with a 7/10 nm thick hydrogen silsesquioxane/HfO$_2$ dielectric layer.  Narrow ($\sim$16 nm) palladium top gates were then deposited, and the electrical resistance measured at 1.6 Kelvin.  Due to the similarity between the top gate width and its distance from the graphene 2DES, the top gate-induced density is well approximated by that of a thin wire above the sample plane and an infinite conducting plane below it:
\be\label{eq:n(x)}
e\rho(x) = \frac{C_{\rm tg}V_{\rm tg}}{1 + x^2/w^2} + C_{\rm bg}V_{\rm bg}
,
\ee
where $C_{\rm tg(bg)}$ and $V_{\rm tg(bg)}$ are the top (bottom) gate capacitance and applied voltage and $w$ is the distance between the wire and the top gate. For such a thin gate, the modulation of the total resistance due to the gate is small; in order to subtract the series resistances of the graphene leads, the numerical derivative of the conductance with respect to the top gate voltage, $dG/dV_{\rm tg}$, was analyzed. 

At zero magnetic field (Fig.\ref{expt}a), $dG/dV_{\rm tg}$ shows distinct behavior in four regions in the ($V_{\rm bg}$, $V_{\rm tg}$) plane, corresponding to ${\rm pp'p}$, ${\rm pnp}$, ${\rm npn}$, and ${\rm nn'n}$ doping, where n (p) refers to negative (positive) charge density and ${}'$ indicates different density. The appearance of FP interference fringes when the polarity of charge carriers in the locally gated region and graphene leads have opposite signs indicates that the mean free path is comparable to the barrier width, $l_{\rm mf}\sim w$.  

In high magnetic field,  in both the bipolar and unipolar regimes, we observe a fan of SdH resonances corresponding to Landau levels (see Fig.\ref{expt}b). 
At lower fields, different behavior is observed depending on the polarity under the gate. In the bipolar regime, as $B$ is lowered, the SdH resonances smoothly evolve into FP resonances. The half-period shift, clearly visible in the data at $B\approx 1\,{\rm T}$, is a hallmark of Klein scattering \cite{Shytov08}. In the unipolar regime, the SdH resonances bend, becoming more horizontal at lower field. The oscillations first begin to lose contrast, and then completely disappear below $B_c\approx 5\,{\rm T}$.

The connection between this behavior and Landau level collapse is exhibited most clearly by a semiclassical analysis. In such analysis, SdH oscillations arise from the interference contribution to the density of states at the Fermi level due to closed trajectories; as a result, Eq.(\ref{eq:BS_quantization_general_2})
with $\epsilon=\epsilon_F$ and $p_y=0$ gives a good estimate for the positions of the transport resonances (see also Fig.\ref{expt}d). For a generic barrier potential, the quantization condition can be written directly in terms of experimental control parameters. Using the Thomas-Fermi approximation, and ignoring the effects of `quantum capacitance' and nonlinear screening \cite{Fogler08}, we define the position-dependent Fermi momentum $k_F(x) = \sqrt{4\pi \rho(x)/g}$, where $g=4$ is the spin/valley degeneracy. Substituting the relation $\epsilon-U(x)=\hbar v_F k_F(x)$ into Eq.(\ref{eq:BS_quantization_general_2}), we obtain
\be\label{eq:BS_quantization} 
\int_{x_1}^{x_2} \sqrt{4\pi \hbar^2 \rho(x)/g - (p_y - eBx)^2}
dx = \pi \hbar n  
.
\ee
We see that the resonance condition is controlled solely by the density profile $\rho(x)$, Eq.(\ref{eq:n(x)}), and is insensitive to the specifics of the single-particle Hamiltonian up to slight modifications due to degree of pseudospin degeneracy and Berry phase. In particular, the quantization condition assumes the same form, apart from the Berry phase contribution,
for massless Dirac particles (monolayer graphene) and massive particles with $g=4$ (bilayer graphene); the spectrum would be only trivially modified for GaAs quantum wells ($g=2$ and $\gamma=0$).

A simple estimate for the critical field can be obtained by comparing the curvature of $\rho(x)$ at $x=0$ with the $B^2x^2$ term in Eq.(\ref{eq:BS_quantization}). Using the device parameters $C_{\rm bg} = 115\,{\rm aF/\mu m^2}$, $V_{\rm bg} = -70\,{\rm V}$, and $w = 50\,{\rm nm}$, we find $B_c = (\hbar/ew)\lp\pi C_{\rm bg}V_{\rm bg}/e\rp^{1/2} \approx 5.2\,{\rm T}$, which matches the data quite well (see Fig.\ref{expt}b). Here we used the relation $C_{\rm bg}V_{\rm bg}+C_{\rm tg}V_{\rm tg}=0$, which defines the boundary at which polarity reversal occurs (white dashed line in Fig.\ref{expt}a).

The actual density profile $\rho(x)$ is nonparabolic, flattening out
on a length scale $2w\approx 100\,{\rm nm}$. However, since $2w$
far exceeds the magnetic length for the fields of interest ($B\gtrsim
1\,{\rm T}$), this flattening does not significantly impact our
discussion of the collapse phenomenon. Indeed, although the states
realized at subcritical magnetic fields are not truly deconfined due to
cyclotron motion in the region outside the top gated region (TGR),
$|x|\gtrsim w$, the corresponding
orbits are very long. For such states, the particle traverses the TGR, makes a partial cyclotron orbit outside of the TGR, and finally
crosses the TGR again to close the orbit (Fig.2a). The net orbit length is a
few $w$, which is much greater than the orbit size at strong fields (i.e. a few
magnetic lengths). The contribution of long orbits to SdH oscillations will be
suppressed due to spatial inhomogeneity and disorder scattering;
hence the distinction between confined and deconfined orbits remains sharp
despite the flattening of the potential.

To estimate the critical field as a function of experimental control parameters $V_{\rm tg}$ and $V_{\rm bg}$ for the density profile (\ref{eq:n(x)}),
we consider an equation for the turning points. Setting $p_y=0$, we have $\hbar k_F(x)=\pm eBx$. Solving this equation and equating the result to barrier half-width, $x_{1 (2)} = \pm w$, we obtain
\be \label{eq:critical_field_from_n(x)}
B_c = (\hbar/ew)\sqrt{(2\pi/e g)(2C_{\rm bg}V_{\rm bg} + C_{\rm tg}V_{\rm tg})}
.
\ee
The observed behavior of $B_c$ is well described by this result (red line in Fig.\ref{expt}b). 

To further explore the effects of curvature, below we analyze the inverted parabola model, $U(x) = -ax^2$. A simple estimate of the collapse threshold can be obtained by considering balance between the Lorentz force and the force due to the electric field, $v_FB=-dU/dx$. This condition is satisfied for a particle moving parallel to the barrier with $x=\ell = ev_FB/(2a)$.
Thus we find an energy-dependent critical field,
\be\label{eq:Bc_quasiclassical}
\label{CritField}B_c(\epsilon) = (2/e v_F)\sqrt{-a\epsilon}
,
\ee
which increases with detuning from neutrality, as in experiment.

A microscopic model of collapse is provided by the Hamiltonian
\be \label{QuantumH} 
H =
\left(\begin{array}{cc} U(x) & v_Fp_-\\
v_Fp_+ & U(x)\end{array}\right) 
,\quad
p_{\pm}=-i\hbar \frac{d}{dx} \pm i(p_y - eBx)
,
\ee
where $p_y$ is the conserved canonical momentum component parallel to the
barrier. We nondimensionalize the problem using ``natural units''
\[
\epsilon_* = (\hbar^2 v_F^2 a)^{1/3},\ \
x_* = \left(\frac{v_F \hbar}{a}\right)^{1/3},\ \
B_* = \frac{\hbar}{e}\left(\frac{a}{v_F \hbar}\right)^{2/3}
.
\]
For each value of $p_y$ and magnetic field $B$, we represent the Hamiltonian as
an $M\times M$ matrix defined on a grid in position space, with periodic
boundary conditions. 
We use the eigenvalues and eigenstates obtained from diagonalization to evaluate the local density of states (DOS) in the middle of the barrier,
\be\label{eq:DOS} N(\epsilon) =
\int\frac{dp_y}{2\pi}\sum_{n=1}^{M}\frac{\gamma}{\pi}\frac{\la
|\psi_{n,p_y}(x=0)|^2\ra}{(\epsilon - \epsilon_n)^2 + \gamma^2} , \ee
with Landau level broadening incorporated through the Lorentzian width $\gamma = 0.2\epsilon_*$. In our simulation, a system of size $L = 15 x_*$ discretized with $M=1500$ points was used.
Averaging with a gaussian weight was used to suppress
the effect of spurious states arising due to a vector potential jump at
the boundary,
\be
\la |\psi_{n,p_y}(x=0)|^2\ra=\int dx' e^{-x'^2/2\sigma^2}|\psi_{n,p_y}(x')|^2
,
\ee
with $\sigma \approx x_*$. Oscillations in the density of states (\ref{eq:DOS}) modulate the rate of electron scattering by disorder, and thus show up in transport quantities measured as a function of experimental control parameters, as in the canonical SdH effect \cite{Abrikosov72}.

The resulting local DOS, shown in Fig.\ref{expt}d, exhibits oscillations which track Landau levels at high $B$. At lower $B$, the behavior is different in the pnp and ${\rm pp'p}$ regimes, realized at positive and negative energies, respectively. In the pnp case, a crossover to FP oscillations is observed. In the pp${}'$p case, discrete Landau
levels give way to a continuous spectrum in the region inside a parabola (dashed line) which marks the quasiclassical collapse threshold,
Eq.(\ref{eq:Bc_quasiclassical}). 

We note that the behavior of $dG/dV_{\rm tg}$ observed in the pnp region, in particular the half-period shift of FP fringes at relatively low $B\lesssim 1T$, is not captured by local DOS, Eq.(\ref{eq:DOS}). As discussed in Ref.\cite{Shytov08}, this half-period shift results from FP interference due to Klein scattering at pn interfaces. A proper model of this effect must account for ballistic conductance in the FP regime.

\begin{figure}
\includegraphics[width=3.25in]{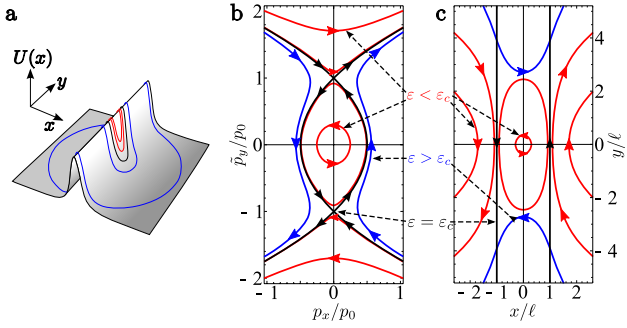}
 \caption[]{
(a) Closed orbits for the Thomas-Fermi potential obtained from the density profile, Eq.(\ref{eq:n(x)}), with $B=9,7,5,3,1\,{\rm T}$
and $p_y=0$. Long trajectories, extending far outside the gated region, do not contribute to SdH oscillations (see text). (b,c) Trajectories for the potential $U(x) = -ax^2$ 
and $p_y = 0$. Three types of trajectories are shown in momentum space (b) and position space (c): subcritical (red), critical (black), and supercritical (blue).
The saddle points in momentum space correspond to motion along straight lines $x = \pm \ell $, where the Lorentz force is balanced by the electric field.
}
\label{figTrajectories}
\end{figure}

The collapse observed in the density of states is related to deconfinement of classical orbits. The orbits can be analyzed as constant energy trajectories of the problem
\be \label{ConstE} \epsilon = v_F \sqrt{p_x^2 + \tilde{p}_y^2} + U(x)
,\quad
\tilde{p}_y = p_y - eBx
.
\ee 
For parabolic $U(x) = -ax^2$ the orbits with $p_y=0$ can be
easily found in polar coordinates $p_x+ip_y=|p|e^{i\theta}$:
%
\be \frac{|p|}{p_0} =
\frac{1}{\sin^2\theta}\left(1 \pm \sqrt{1
-\frac{\epsilon}{\epsilon_c}\sin^2\theta}\right)
, \quad
\epsilon_c =\frac{(v_FeB)^2}{4a}
\label{p-traj} 
\ee 
with $p_0 = v_F e^2B^2/2a$ 
(see
Fig.\ref{figTrajectories}b). Only real, positive solutions should be retained; when $\epsilon/\epsilon_c > 1$, the discriminant in Eq.(\ref{p-traj}) is negative near $\theta \approx \pi/2$ and trajectories cannot close (blue curves in Fig.\ref{figTrajectories}b).

The related orbits in position space can be found from the relation $dy/dx=\dot y/\dot x=\tilde p_y/p_x$, giving
\be \label{dy/dx} 
\frac{dy}{dx} = \frac{\pm v_F(p_y - eB x)}{\sqrt{(\epsilon - U(x))^2
- v_F^2(p_y - eB x)^2}}.
\ee
For $p_y = 0$, integration is 
performed using the variable $u = x^2/\ell ^2$,
\be \frac{y}{\ell } = \pm\int\frac{du}{\sqrt{(u + \epsilon/\epsilon_c - 2)^2 +
4(\epsilon/\epsilon_c - 1)}} 
,
\ee
where $\ell  = v_F eB/2a$ such that $\epsilon_c = a\ell ^2$.
The integrand 
changes its behavior at the critical energy $\epsilon_c$.
For $\epsilon > \epsilon_c$, the integrand is real valued for all $u$ and 
\be \sinh\lp \frac{y(x) - y_0}{\ell }\rp = \frac{x^2/\ell ^2 +
\epsilon/\epsilon_c - 2}{2\sqrt{\epsilon/\epsilon_c - 1}}. 
\label{RS1} 
\ee
For $\epsilon < \epsilon_c$, real solutions are divided into two domains $0 \le
u \le 2 - \epsilon/\epsilon_c - 2\sqrt{1-\epsilon/\epsilon_c}$ (closed
orbits)  and $u > 2 - \epsilon/\epsilon_c + 2\sqrt{1 - \epsilon/\epsilon_c}$
(open orbits): 
\be \label{RS2} 
\cosh\lp \frac{y(x) - y_0}{\ell }\rp=\pm
\frac{2-\epsilon/\epsilon_c -x^2/\ell ^2}{2\sqrt{1 - \epsilon/\epsilon_c}}
.
\ee
The red curves in Fig.\ref{figTrajectories}c correspond to the low energy
regime, $\epsilon<\epsilon_c$, where orbits can either be closed (Landau levels) or open (trajectories for particle moving far from the barrier). At higher energies, $\epsilon>\epsilon_c$, all trajectories are open. The
straight black lines correspond to the critical orbits of Eq.(\ref{CritField}),
where the Lorentz force and electric field are balanced. In addition to the two
particular critical trajectories shown, in the limit $\epsilon/\epsilon_c
\rightarrow 1$ there is an entire family of critical trajectories which
asymptotically approach these lines.

Interestingly, unlike in the case of the potential obtained from the Thomas-Fermi model, where the
classical turning points move continuously to infinity as the transition is
approached, trajectories in the parabolic potential are trapped between the
critical separatrix lines. At very low energies, closed orbits are
approximately circular; as the energy increases towards $\epsilon_c$, orbits become more and more elongated, until finally merging with the separatrix at $\epsilon=\epsilon_c$ (see Fig.\ref{figTrajectories}).

In summary, graphene devices with a barrier induced by a narrow top gate can be used to probe electronic states on the spatial scale of a few tens of nanometers. In our transport measurments, the SdH-type resonances arising from quantized states associated with closed orbits are used to directly observe the competition between magnetic confinement and deconfinement due to electric field. As a result of this competition, the discrete spectrum of Landau levels collapses when subjected to a strong external potential. Experimental observations are found to be in good agreement with theory.

MSR acknowledges NSF support under DMR090647 and PHY0646094. PK acknowledges support from NRI, INDEX, ONR MURI, and FENA. LSL acknowledges ONR support under N00014-09-1-0724.


\end{document}